# Statistical Design of Pragmatic Trials Using Electronic Health Record Data when Outcome Assessments are Uncontrolled and Irregular

**Authors**

Jennifer F. Bobb, PhD[1,3]

Sungtaek Son, MS[3]

Melissa L. Anderson, MS[1]

Noorie Hyun, PhD[1,3]

Lynn L. DeBar, PhD, MPH[2]

Katharine A. Bradley, MD[1]

**Affiliations**:

[1]Kaiser Permanente Washington Health Research Institute; Seattle, Washington, USA

[2]Kaiser Permanente Center for Health Research Northwest Region; Portland, Oregon, USA

[3]Department of Biostatistics, School of Public Health, University of Washington; Seattle, Washington, USA

**Corresponding Author:** Jennifer F. Bobb

Kaiser Permanente Washington Health Research Institute, 1730 Minor Ave., Ste 1360, Seattle, WA 98101
Jennifer.F.Bobb@kp.org

**Abstract**

**Background:** Pragmatic trials increasingly define outcomes using real-world data such as electronic health records (EHRs), where assessments are collected during routine care rather than at fixed study-specific timepoints. Consequently, these uncontrolled assessments may be irregular, sparse at timepoints of interest, and in some trials, affected by the intervention itself. We refer to this latter setting as *intervention-dependent assessments*. Because assessments occur after randomization, intervention-related differences in the outcome measurement process can lead to biased treatment effect estimates.

**Methods:** We developed a simulation study framework to inform the choice of statistical method for trials with uncontrolled, potentially intervention-dependent assessments, which we applied to the MI-CARE pragmatic trial. We first constructed a pre-trial, retrospective-cohort dataset, which mimicked the trial's EHR-based eligibility criteria and outcome assessment process. We then combined estimates of assessment frequency and timing with scientific assumptions on how the intervention might alter these distributions to simulate data under sparse and intervention-dependent assessments. Comparator approaches included simple methods that select a single follow-up measure per person (e.g., best or randomly selected), and longitudinal models using all available follow-up scores. Methods were compared to identify the most powerful among unbiased methods.

**Results:** Under intervention-dependent assessments, naïve methods such as using the "best" score or using a randomly selected score without adjusting for measurement timing produced substantial bias. Models that adjusted flexibly for the follow-up timing estimated time-point specific or time-averaged treatment effects without bias. Simulation results informed the selection of the statistical approach for the MI-CARE trial. Among the methods considered, the most powerful, unbiased approach was a linear mixed model with exponential correlation structure, adjustment for time since baseline, and a time-varying intervention effect (modeled flexibly via splines) to estimate the intervention effect at 12 months (end of the intervention window).

**Conclusions:** Pre-trial data can be used to conduct a simulation study tailored to the trial's data features (e.g., assessment frequency/sparsity) to inform the choice of estimand and analytic approach to maximize the study's power. Trials with uncontrolled assessments should consider the potential for intervention-dependent assessments and select an appropriate method to avoid bias.

**Trial registration:** ClinicalTrials.gov identifier NCT05122676

# 1. Background

Pragmatic trials are increasingly being conducted that use real-world data such as electronic health records (EHRs) to define study outcomes. [1–5] Using data that is routinely collected as part of clinical care, rather than collecting new, study-specific primary data, enables trials to be conducted efficiently within large study populations (e.g., entire clinical practices) and enhances generalizability of study findings.[4–6] In addition, innovative designs can be employed, such as encouragement or Zelen designs, in which patients are randomized either to be offered the intervention or to a control group. In these designs, real-world data allow outcomes to be obtained for all patients, including those randomized to the intervention who do not consent and those randomized to control who are never contacted.

However, unlike traditional trials in which patients consent and outcomes are collected at pre-specified follow-up time points controlled by the study, pragmatic trials conducted in more naturalistic settings rely on existing data from routine care documented in insurance claims or EHRs for study outcomes. This use of existing data results in uncontrolled outcome assessments, in which the study team does not control the measurement process (including timings of when measurements are collected), leading to distinct challenges for trial design and analysis.

First, uncontrolled outcome assessments result in irregular observation times and a variable number of follow-up measures across participants. Studies must have a well-defined time point for the primary outcome,[7] but when follow-up times are irregular, power for estimating the intervention effect at a particular point in time could be limited. Potential analytic methods include simple approaches, such as selecting a single follow-up measure per person (e.g., random measure, or measure closest to a desired time point), or longitudinal methods that use all follow-up measures, such as generalized linear mixed models (GLMM) or generalized estimating equations (GEE).[8,9] Such approaches may estimate different quantities (i.e., *estimands*)[10,11] and have different statistical properties, including efficiency (e.g., narrower confidence intervals) when measures are irregular across participants.

Second, uncontrolled and irregular observation times could be affected by *informative assessments*, whereby the presence of an observed outcome at a particular measurement time could be impacted by variables (measured or unmeasured) including baseline patient characteristics, or time-varying factors after randomization.[9,12–27] An example of informative assessments occurs when the measurement process is associated with the outcome of interest, for example if patients are followed-up more regularly due to health concerns related to the outcome (e.g., depression severity). Several statistical methods are available that can provide valid (i.e., unbiased) intervention effect estimates under different assumptions on the mechanism for the assessment process.[9]

In this paper, we focus on a particular form of informative assessments in which the intervention being tested affects the outcome measurement process, including the number and timing of measures documented during the follow-up period. We refer to this setting as *intervention-dependent assessment*s*.* For example, an intervention involving measurement-based care[28,29] may prompt clinicians to collect patient-reported outcomes as part of routine care and document these measures in the EHR, thereby increasing outcome measurement in the intervention arm. Although randomization ensures balance with respect to baseline factors, assessment times occur after randomization and therefore may not be balanced across trial arms. Consequently, differences in the measurement process across arms could lead to biased intervention effect estimates if not handled appropriately. A prior study considered a general setting in which the outcome assessment process depends on a covariate, showing through theory and simulation that modeling the intervention effect over time can avoid bias.[30] This setting encompasses intervention-dependent assessments, where the covariate influencing the assessment process is the intervention group itself. However, this prior work focused primarily on bias and did not address how to choose among candidate unbiased methods based on statistical efficiency or power. Moreover, they did not consider settings with highly irregular and sparse assessments, in which many individuals may not have repeated measures over time and power may vary substantially across analytic approaches. Thus, a comprehensive simulation, that generates a realistic assessment process based on real-world data and that evaluates both bias and efficiency can inform the choice of analytic approach for trials with uncontrolled, irregular, and potentially intervention-dependent assessments.

The MI-CARE trial is a pragmatic encouragement trial, conducted among primary care patients with opioid use disorder (OUD) and depression, to evaluate a collaborative care model for treating co-occurring conditions.[31] A key study outcome is the change in depressive symptoms, as measured by the Patient Health Questionnaire (PHQ-9) score documented in the EHR. Though this measure is collected as part of routine care, because the embedded intervention includes measurement-based care, it is expected to increase measurement of PHQ-9s relative to usual care. In addition, pre-trial data from MI-CARE showed that assessments are highly irregular, with considerable variability across participants (**Figure 1**). Among those having at least one PHQ-9 follow-up score, 36.9% had just a single score, whereas 24.5% had ≥5 scores (up to a maximum of 40 scores). Moreover, though a clinically meaningful endpoint for capturing the primary outcome is at 12 months (end of the intervention period), a high proportion of individuals did not have a follow-up score proximal to this time point: among individuals with at least one PHQ-9 follow-up score, only 23% had a measure in month 12 (and 49.8% had a measure within a broader 10-13 month window). This variation in availability of follow-up data across individuals highlights the challenges in selecting an analytic approach. Using data from a single time point (e.g., requiring a measure within the 10-13 month window when 12-month outcomes are of interest) would result in >50% missingness, limiting power and leading to increased potential for missing data bias. On the other hand, longitudinal analyses can leverage all follow-up scores, but repeated values are not available for over one third of individuals. Consequently, the choice of analytic approach that avoids bias due to intervention-dependent assessments while maximizing power given the highly irregular assessment process is unclear. We therefore sought to conduct a simulation study to inform the statistical analysis plan.

In this work, we illustrate a simulation study framework for selecting the statistical analysis approach in the presence of uncontrolled assessments, which we apply to the MI-CARE pragmatic trial. Our simulation study incorporates estimates on the anticipated degree of data irregularly/sparsity from pre-trial data that mimics trial eligibility criteria, as well as a range of scientific assumptions on how the intervention might alter these distributions. We evaluated candidate analytic approaches to identify an analytic approach for the MI-CARE trial that avoids bias due to intervention-dependent assessments while maximizing power given expected data sparsity.

# 2. Methods

This case study illustrates how the statistical study design of a trial with highly irregular, and anticipated intervention-dependent, assessments can be informed by conducting a simulation study using pre-trial data. MI-CARE is a pragmatic, individually randomized encouragement (Zelen) trial[32,33] evaluating whether a collaborative care intervention is effective for treating primary care patients with both OUD and depression relative to usual care, conducted in two health systems.[31] This case study explores the choice of analytic approach for the primary depression outcome (change in PHQ-9 score from baseline to follow-up) documented in the EHR over the 3-13 month follow-up window. Specifically, we sought to select an analytic approach for estimating the intervention effect that was unbiased in the presence of intervention-dependent assessments. We further sought to identify the most powerful (from among unbiased methods) given the highly variable number of measures across participants and sparse data availability at time points of interest. To ensure our simulation study would be relevant to our trial setting with highly irregular assessments, we first developed a retrospective cohort dataset, which used pre-trial data that mimicked trial eligibility criteria, and then used plasmode simulation methods[34] to resample observations from this cohort.

### *2.1. Data generation*

Data generation was informed by the design of the MI-CARE trial (e.g., sample size, stratified randomization), real-world data from our retrospective cohort reflecting usual care, and clinical co-investigator input on the how the intervention was expected to alter distributions relative to usual care. First, we randomly assigned patients (n=800; 400 in each health system) with a 1:1 allocation to intervention and control arms stratified by health system. We then generated the number of follow-up scores and their timing (i.e., days since baseline) for each person depending on the intervention group status. For usual care patients, we used the empirical distribution of the number and timing of follow-up PHQ-9 scores by resampling observations from our retrospective cohort dataset. For intervention patients we considered two different scenarios: (1) an "optimal" scenario in which patients have a follow-up score monthly during the 3-13 month follow-up window, with the day within each month randomly selected; and (2) a more realistic scenario in which 20% of patients have optimal follow-up, 60% have follow-up for the first month of the 3-13 month follow-up window, due to anticipated early engagement and drop out over time (with the remaining months' follow up the same as usual care), and 20% having follow-up as under usual care. **Figure 2A** illustrates the distribution of follow-up times under these data generating scenarios. The mean (SD) number of follow-up scores across participants was 3.2 (3.89) under usual care and 10.0 (0.00) and 4.9 (4.33) under the optimal and realistic follow-up scenario, respectively.

We then generated outcome values from a linear mixed model (LMM) that accounted for correlation of repeated scores within a person using an exponential covariance (structure with the lowest Akaike information criterion [AIC][35] in the retrospective cohort). The mean model was of the form

$$\mu_{ij} \coloneqq E\left[Y_{ij}\right] = \alpha + \beta_0(t_{ij}) + \beta_1 baseline_i + \beta_2 site_i + \Delta(t_{ij}) \cdot A_i$$

where $Y_{ij}$ is outcome $j$ for person $i$ at follow-up time point $t_{ij}$, $baseline_i$ is the baseline score, $site_i$ is the health system, and $A_i$ is the randomized intervention group. The term $\beta_0(t_{ij})$ represents the trend in the outcome over time (under usual care), and $\Delta(t_{ij})$ represents the time-varying intervention effect. The parameters used in the data generating model were estimated from fitting the above model (with no intervention effect) to the retrospective cohort data. Though retrospective cohort data suggested little evidence of a time-varying trend $\beta_0(t)$ in the outcome, we considered multiple scenarios including linear and quadratic time trends (**Figure 2B**). We considered three scenarios for $\Delta(t_{ij})$: no effect, constant effect over time, and a time-varying effect reflecting the hypothesis that depression symptoms would decline with increasing engagement with the intervention over the 12-month intervention window (**Figure 2C**).

## *2.2. Candidate Analytic Approaches*

All approaches used linear models and adjusted for health system and the baseline score. Given the meaningful proportion of patients with only a single follow-up score (**Figure 1**), we considered both methods that use a single summary of the follow-up PHQ-9 scores (randomly selected score, score closest to 12 months [end of the intervention window], mean score, and best [lowest] score), as well as longitudinal methods that use all available scores: GEE weighted to account for the potential for informative cluster size (observations weighted by the inverse of the number of follow-up measures $n_i$) and with working independence covariance structure ("weighted independence estimating equations");[36] and LMM with different assumed correlation structures (exchangeable, continuous autoregressive [CAR], exponential [Exp]). For LMM, we did not formally account for informative cluster sizes, because previous work suggests that for random intercept models (such as the ones considered here) ignoring informative cluster sizes results in little bias in estimating covariate effects.[16]

For each method, we considered versions that did and did not adjust for time since baseline of the follow-up scores. For the method using the mean follow-up score, we additionally considered models that were adjusted or weighted by the number of follow-up scores. When adjusting for time since baseline, we considered linear adjustment, as well as flexible adjustment using natural cubic splines with 3 degrees of freedom (DF) and knots at quantiles. We additionally considered versions that (1) estimated a constant intervention effect over time or (2) allowed the intervention effect to vary over time (modeled using splines with 3 DF) and estimated the effect at 12 months. Standard errors (SEs) were calculated using robust (sandwich) variance estimators for methods selecting a single score per person and for the GEE methods using all available scores; SEs for LMM used model-based estimates.

Under each data-generating scenario, we generated 5000 datasets and applied the analysis approaches listed above. We estimated bias of the estimated intervention effect relative to the target estimand (e.g., average effect or effect at 12 months), as well as the type 1 error rate (when $\Delta(t_{ij}) = 0$) or power (when $\Delta(t_{ij}) \neq 0$).

# 3. Results

Given the large number of 24 possible simulation data-generating scenarios, we present results from combinations representing important scenarios of interest. We first focus on scenarios with no intervention effect, comparing methods that summarize the follow-up scores into a scalar summary (**Table S1** in the Supplement). We found that using the best score yielded poor performance, due to severe bias and type 1 error inflation; using the mean score also performed poorly, with estimated standard error (SE) underestimating the empirical SE, resulting in higher type 1 error, even with adjustment or weighting by the number of follow-up scores. We therefore discarded these two methods for the remaining simulations. On the other hand, using a randomly selected score or the score closest to 12 months resulted in unbiased estimates and correct type 1 error rates, provided the model adjusted for time since baseline of the selected score with sufficient flexibility (**Table S2**). Models that did not adjust for time since baseline resulted in moderate to severe bias (and type 1 error inflation) when the distribution of assessment times differed between arms (**Figure 2A**). We therefore focused on models with flexible adjustment for time since baseline in subsequent scenarios.

Next, we compared methods that select a single score to methods that use all follow-up measures, focusing on the realistic follow-up scenario (**Table 1**). Under no intervention effect, GEE and the LMM with the exponential or CAR correlation structures were unbiased with nominal type 1 error; the LMM with the exchangeable correlation had a slightly elevated type 1 error under the spline model (due to the SE underestimating the empirical SE).

Under a constant intervention effect, all methods were unbiased. Among models that assume a constant effect, methods selecting a single score only had slightly reduced power to estimate this constant effect relative to models using all scores (97% for random score and score closest to 12 months vs 99% for GEE and LMM). For models that allow for a time-varying effect (when estimating the effect at 12 months), power is reduced to 92% for LMM and 73% for GEE due to the greater variability in the weighted GEE estimates. Because our data-generating model was an LMM, this performance may be optimistic, though similar performance under misspecification of the covariance (CAR versus Exp structure) suggests some robustness (**Table S3** in the Supplement). Selecting the score closest to 12 months but allowing for a time-varying effect surprisingly led to higher power than GEE at 79%. This may be due to the different distribution of the follow-up score timings under these two methods (**Figure 2A**), such that applying splines with the same degree of flexibility leads to slightly more precision for estimating the score at 12 months when the distribution of follow-up scores is more highly concentrated at 12 months.

For scenarios assuming a time-varying intervention effect, we focus our summary on power of competing methods, as comparing estimates from these methods is less valuable due to their estimating different quantities (e.g., the effect at 12 months versus a constant, time-averaged effect). For models estimating a constant effect, using the score closest to 12 months had the highest power at 71%, followed by the LMM at ~40%, and then GEE at 35% and selecting a random score at <30%. However, this was likely due to our assumed time-varying $\Delta(t)$ that increased in magnitude over time; if instead the magnitude were highest at 3 months and then subsequently decreased, the method selecting the score closest to 12 months would not have the highest power. For models that allow for a time-varying effect, when estimating the effect at 12 months LMM have the highest power (~91%) followed by using the score closest to 12 months (79%), GEE (75%), and using a random score (~50%).

### *3.1. Implications for MI-CARE*

During protocol development, given expected issues with using the “best” score per person (due to potential for higher frequency of follow-up under the intervention) and sparsity of follow-up measures, the preliminary proposed approach was to use the score closest to 12 months (end of the intervention window), adjusted for the timing of the follow-up score. Through conducting this work, we clarified that–under the reasonable assumption that there is likely to be a time-varying intervention effect–the initial analytic approach of selecting the score closest to 12 months results in a different estimand as compared to fitting a model that allows for a time-varying intervention effect and estimating the score at 12 months.[1] Under the time-varying effect that we considered based on scientific input, and given the anticipated degree of irregularity in assessment times based on our retrospective cohort, using a LMM had the highest power and yielded unbiased intervention effect estimates. Our team therefore decided to use a LMM (with exponential covariance) to estimate the time-varying intervention effect $\Delta(t)$, with the primary outcome defined as the intervention effect at 12 months.

## 4. Discussion

This work sheds light on how uncontrolled outcome assessments can pose threats to the validity of pragmatic trials using EHR data for study outcomes. Through a comprehensive simulation study motivated by a real-world trial, in which an embedded intervention could impact timing and frequency of measurement of patient-reported outcomes in the EHR, we examined alternate analytic approaches for handling irregular observation times. With intervention-dependent assessments, naïve methods such as using the “best” score over the follow-up period or using a randomly selected score without adjusting for time since baseline can lead to severe bias.

---

[1] For models that include a main effect for the intervention and adjust for time flexibly, in the presence of a true time-varying effect, estimators of the intervention effect based on these models corresponds to a weighted treatment effect estimand, even when the model is mis-specified (e.g., due to true interaction of $A_i$ with $t$ corresponding to a time-varying treatment effect). The weights relate to the overlap in the distribution of follow-up score timings among those observations included in the analysis: observations are given higher weights if their follow-up time has larger overlap between intervention and control groups.[37]

Importantly, different approaches estimate different estimands, including time-point specific and (weighted) time-averaged treatment effect estimates. We further illustrated how retrospective cohort data, collected prior to the trial but which mimic trial eligibility criteria, can be used to design a realistic simulation to inform the choice of analytic approach tailored to the specific trial setting (e.g., degree of data sparsity for the outcome of interest).

Although the importance of selecting an estimand prior to determining the analytic approach has been emphasized in both prior statistical literature[38,39] and in guidance from regulatory agencies,[11] in practice approaches targeting different estimands yield different statistical properties including efficiency and power, as demonstrated by our simulation study. For example, if observations are sparse at a particular time point of interest, power could be low for the desired estimand and alternates may need to be considered, such as the effect at a different time point or averaged over (part of) the follow-up window. Conducting a power analysis informed by pre-trial data and guided by realistic assumptions on how the intervention may alter distributions—of both the assessment process and outcome values—can be used to guide both the choice of estimand and analytic approach.

Several expansions of our simulation study could be considered. First, alternate correlation structures could be used to generate outcome data. We found that the LMM with the exponential correlation structure performed the best, which is likely due in part to the use of this distribution for the data generating model. However, the CAR correlation structure, despite being mis-specified, performed similarly well, and a sensitivity simulation generating data from an unstructured covariance resulted in similar performance (**Table S3**) suggesting that this finding is robust. Second, the GEE approach we considered accounted for informative cluster sizes,[36] though we did not formally generate outcome data in which informative cluster sizes occurred (e.g., severity of the outcome did not depend on the number of measures in our simulation). Future studies could examine whether incorporating informative cluster sizes in data generation might impact the relative performance of competing methods.

Additional future work could consider data-adaptive methods to select the degree of smoothing for modeling time-varying trends, rather than select the DF a priori as we have done. Moreover, our focus was on intervention-dependent assessments, whereas future work could incorporate outcome-dependent assessments (e.g., when the missing values depend on the unobserved outcomes).[9,19,40] Several methods have been developed to address other settings of informative assessments including outcome-dependent assessments, but examination of their relative performance in settings with a highly irregular and sparse outcome assessment process is warranted. Lastly, for simplicity our simulation did not account for individuals expected not to have any follow-up scores (up to 40% based on preliminary data), because we sought to focus on the choice of methods to address the irregular (rather than completely missing) assessment process. In our statistical analysis plan for MI-CARE, we plan to address this missingness by using a covariate-adjusted LMM adjusted for baseline factors associated with missingness or with the outcome, and to consider sensitivity analyses to account for the possibility of outcome-dependent assessments.[18]

## 5. Conclusion

Future pragmatic trials with irregular assessments should consider the potential for intervention-dependent assessments and select an appropriate analytic method to avoid bias. Furthermore, if possible given trial timelines, we recommend using real-world data prior to the trial that mimic trial eligibility criteria to develop a simulation study tailored to the particular data features of the study (e.g., variation in the number and sparsity of assessments across individuals) to inform the choice of analytic approach to maximize the study's power.

## Abbreviations

EHR: Electronic health records
GLMM: Generalized linear mixed models
GEE: Generalized estimating equations
OUD: Opioid use disorder
PHQ-9: Patient Health Questionnaire
LMM: Linear mixed model
AIC: Akaike information criterion
CAR: Continuous autoregressive
Exp: Exponential
DF: Degrees of freedom
SE: Standard error

## Declarations

**Ethics approval and consent to participate**:

Advarra provided institutional review board (IRB) oversight of the study (Protocol No. 00045256) and provided waivers of consent and HIPAA authorization for sample identification, baseline data, randomization and outcome measures for the trial. Local IRBs at each study site ceded local authority to Advarra and reviewed for local context where appropriate.

**Competing interests**:

The authors declare that they have no competing interests

**Funding**:

This research was supported by a grant to Drs. DeBar and Bradley (Multiple PIs) from the National Institute of Mental Health (UF1MH121949). The content of this publication is solely the responsibility of the authors and not necessarily of the National Institute of Mental Health, the National Institutes of Health, or the United States federal government.

**Authors' contributions**:

JFB conceived of this study. All authors contributed to the study design. MA performed data cleaning. SS performed the analyses. JFB wrote the initial draft of the manuscript. All authors contributed to interpreting findings and critical revisions of the manuscript and approved the final manuscript.

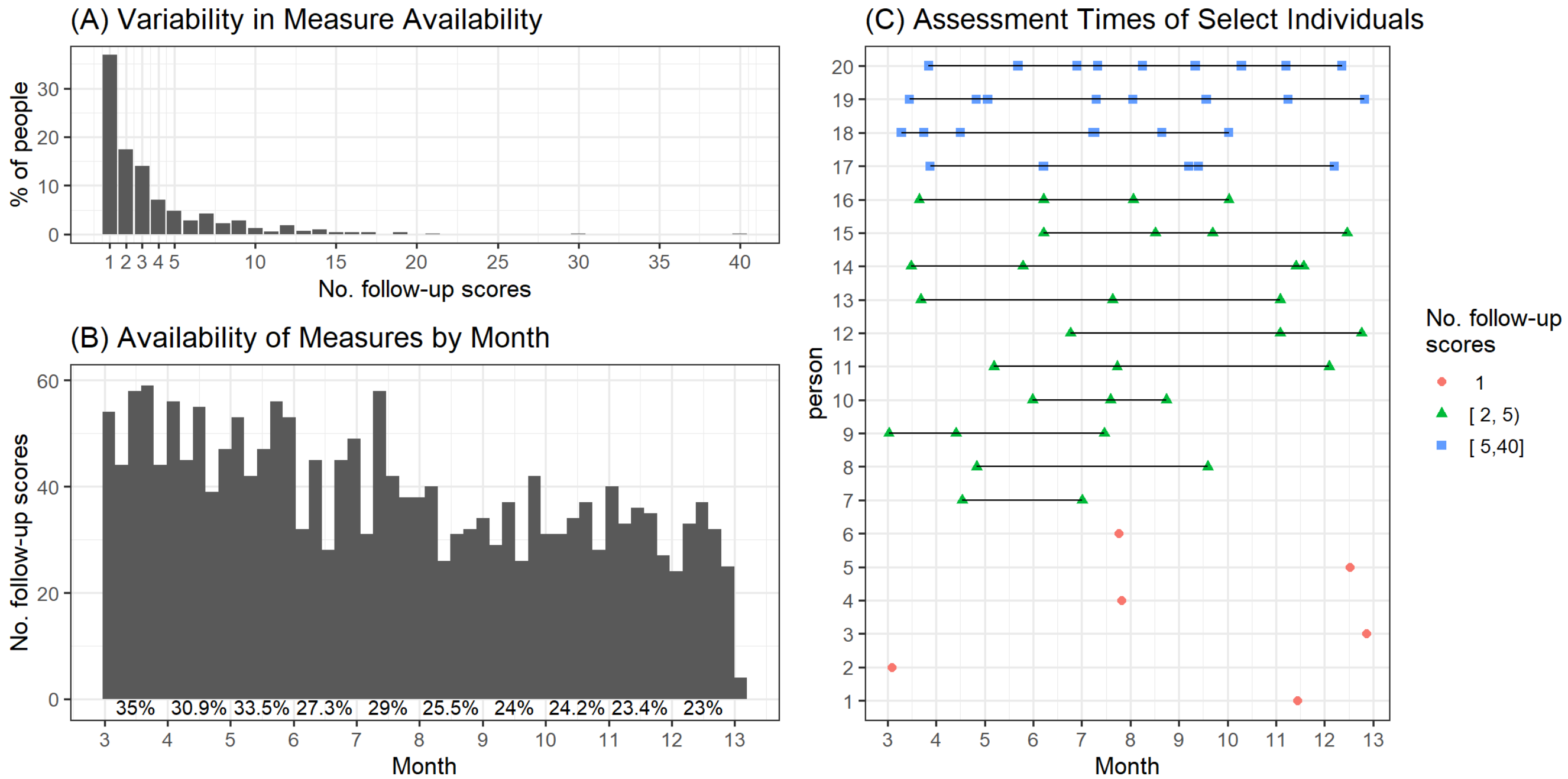

**Figure 1.**

Title: Frequency and timing of PHQ-9 outcomes from retrospective cohort data prior to the MI-CARE trial

Legend: Summaries shown among individuals with at least 1 follow-up measure. In Panel (B) the percentages (%) at the bottom of the plot indicate the % of individuals who have at least one follow-up measure within the given month (note %s do not sum to 100 since individuals can have measures in multiple months).

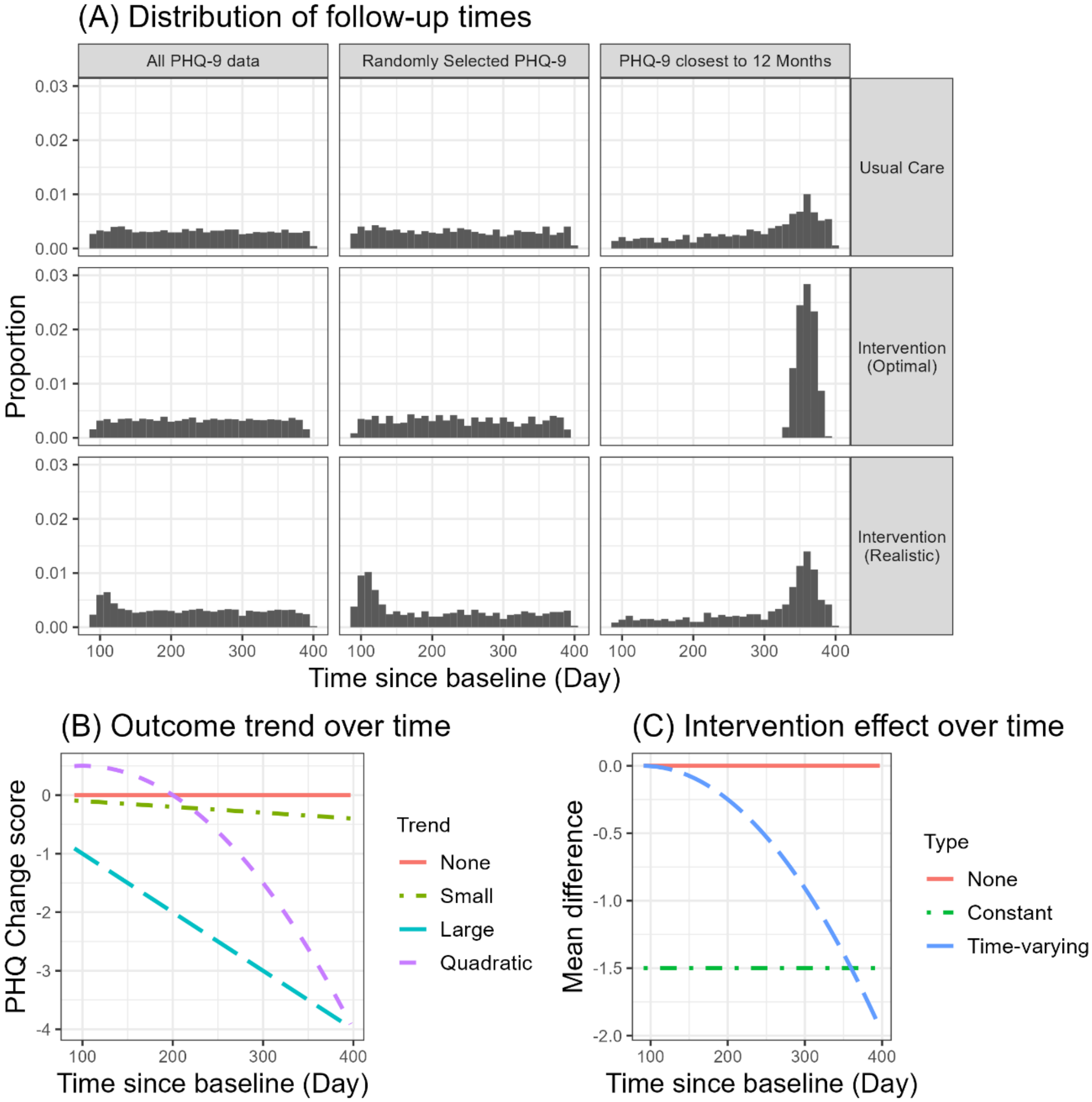

**Figure 2.**

Title: Data generating scenarios for the simulation study

Legend: Simulation study examines the choice of analytic method for the depression outcome of change in PHQ-9 score from baseline to follow-up. (A) Distribution of follow-up times (y-axis denotes proportion) for usual care and for the intervention group under the optimistic and realistic scenarios of follow-up using all PHQ-9 data during follow-up (left column ["All"]), a randomly selected PHQ-9 during follow-up (middle column ["Random"]), and the PHQ-9 closest to 12 months (right column ["12th month"]). (B) Scenarios of assumed outcome trend over time since baseline, $\beta_0(t)$. (C) Scenarios of assumed intervention effect over time, $\Delta(t_{ij})$.

**Table 1.** Simulation study results comparing methods across scenarios of intervention effect

| Modeling and Estimation Approach | | Operating characteristics | | | | |
|---|---|---|---|---|---|---|
| Choice of Follow-up Score | Modeling time-varying intervention effect* | Estimate | Empirical SE | SE (median) | Coverage (95% CI) | Rejection Rate |
| *(A) No intervention effect* ($\Delta \coloneqq \Delta(12) = 0$) | | | | | | Type 1 error |
| Random score | none | 0.009 | 0.393 | 0.392 | 0.949 | 0.051 |
| | splines (3 DF) | -0.013 | 0.805 | 0.780 | 0.939 | 0.061 |
| Score closest to 12 months | none | 0.009 | 0.389 | 0.390 | 0.948 | 0.052 |
| | splines (3 DF) | -0.006 | 0.548 | 0.545 | 0.944 | 0.056 |
| Weighted GEE | none | 0.001 | 0.351 | 0.353 | 0.952 | 0.048 |
| | splines (3 DF) | -0.013 | 0.580 | 0.572 | 0.948 | 0.052 |
| LMM / Exchangeable | none | -0.007 | 0.349 | 0.351 | 0.951 | 0.049 |
| | splines (3 DF) | 0.002 | 0.468 | 0.419 | 0.923 | 0.077 |
| LMM / CAR | none | 0.001 | 0.354 | 0.350 | 0.951 | 0.049 |
| | splines (3 DF) | 0.007 | 0.449 | 0.447 | 0.948 | 0.052 |
| LMM / Exp | none | -0.006 | 0.353 | 0.350 | 0.947 | 0.053 |
| | splines (3 DF) | 0.002 | 0.444 | 0.447 | 0.954 | 0.046 |
| *(B) Constant intervention effect* ($\Delta \coloneqq \Delta(12) = -1.5$) | | | | | | Power |
| Random score | none | -1.499 | 0.390 | 0.393 | 0.951 | 0.969 |
| | splines (3 DF) | -1.507 | 0.779 | 0.779 | 0.951 | 0.487 |
| Score closest to 12 months | none | -1.495 | 0.387 | 0.390 | 0.949 | 0.966 |
| | splines (3 DF) | -1.506 | 0.549 | 0.544 | 0.944 | 0.788 |
| Weighted GEE | none | -1.501 | 0.354 | 0.353 | 0.947 | 0.987 |
| | splines (3 DF) | -1.506 | 0.582 | 0.573 | 0.947 | 0.734 |
| LMM / Exchangeable | none | -1.498 | 0.354 | 0.350 | 0.946 | 0.988 |
| | splines (3 DF) | -1.495 | 0.473 | 0.420 | 0.916 | 0.918 |
| LMM / CAR | none | -1.499 | 0.354 | 0.350 | 0.948 | 0.990 |
| | splines (3 DF) | -1.502 | 0.442 | 0.447 | 0.955 | 0.922 |
| LMM / Exp | none | -1.496 | 0.341 | 0.350 | 0.955 | 0.994 |
| | splines (3 DF) | -1.493 | 0.444 | 0.447 | 0.952 | 0.917 |
| *(C) Time-varying intervention effect* ($\Delta(12) = -1.5$)** | | | | | | Power |
| Random Score | none** | -0.554 | 0.400 | 0.393 | 0.944 | 0.298 |
| | splines (3 DF) | -1.510 | 0.770 | 0.778 | 0.950 | 0.494 |
| Score closest to 12 months | none** | -0.987 | 0.391 | 0.390 | 0.949 | 0.715 |
| | splines (3 DF) | -1.510 | 0.552 | 0.544 | 0.945 | 0.792 |
| Weighted GEE / Independence | none** | -0.566 | 0.357 | 0.354 | ** | 0.349 |
| | splines (3 DF) | -1.507 | 0.582 | 0.571 | 0.946 | 0.749 |
| LMM / Exchangeable | none** | -0.588 | 0.348 | 0.351 | ** | 0.395 |
| | splines (3 DF) | -1.501 | 0.473 | 0.420 | 0.918 | 0.922 |
| LMM / CAR | none** | -0.595 | 0.352 | 0.350 | ** | 0.400 |
| | splines (3 DF) | -1.505 | 0.461 | 0.448 | 0.944 | 0.910 |
| LMM / EXP | none** | -0.582 | 0.352 | 0.350 | ** | 0.390 |
| | splines (3 DF) | -1.506 | 0.454 | 0.447 | 0.949 | 0.913 |

Results compare approaches that select a single follow-up score per person or that use all scores, under different scenarios of the intervention effect over time. The simulation setting presented is the realistic scenario of follow-up for intervention participants, and assuming a small, linear outcome trend over time.

* Models estimating a time-varying effect report the estimated intervention effect at 12 months.

**Models estimating a constant effect are estimating a different estimand (i.e., treatment effect) than the effect at 12 months. Under a time-varying intervention effect scenario, the value of this estimand depends on the distribution of the follow-up times and was empirically calculated as -0.5374 and -0.997 when using the distribution of a randomly selected score and the score closest to 12 months, respectively. When using all scores, the true value of this estimand is not easily computable; we therefore do not present coverage.

## Supplement

### *Simulation study results comparing methods selecting a single score per person*

In this section we present results for methods that summarize the follow-up scores into a scalar summary (e.g., by selecting a single score or using the mean score). For this set of simulations, we focus on the scenario that assumed no intervention effect, no trend in the outcome over time, and optimal follow-up among the intervention group, to highlight the effect of different assessment time distributions (**Table S1**). We see that using the best (lowest) follow-up score performs very poorly as expected given the greater number of follow-up scores among intervention patients, which is not ameliorated by adjustment for the number of scores. Using the mean follow-up score results in only minimal bias but the SEs—even when applying weighting—underestimate the empirical (Monte Carlo) SE, resulting in elevated type 1 error rates. Using a randomly selected score or the score closest to 12 months yielded unbiased estimates and correct type 1 error rates. We note that SE estimates are larger when selecting the score closest to 12 months and adjusting for the follow-up timing, because in this setting the timing of the follow-up score is related to the intervention group but not with the outcome (since this scenario assumes no time trend in the outcome). Given the poor performance using the mean or best follow-up score, these methods were discarded for the remaining simulation scenarios.

**Table S1.** Characteristics of approaches that select a single score per person or use a scalar summary of the scores, under the scenario with optimal follow-up for intervention participants, no intervention effect, and no trend in the outcome over time

| Modeling and Estimation Approach | | Operating characteristics | | | |
|---|---|---|---|---|---|
| Choice of Follow-up Score | Adjustment for time since baseline or number of scores | Bias | Empirical SE | SE (median) | Type 1 error |
| Mean score | Unadjusted, weighted* | -0.015 | 0.547 | 0.462 | 0.092 |
| | Adjusted (no. scores) | -0.016 | 0.712 | 0.633 | 0.086 |
| | Adjusted (no. scores), weighted* | -0.018 | 0.712 | 0.636 | 0.082 |
| Best (lowest) score | Unadjusted | -3.190 | 0.368 | 0.364 | 1.000 |
| | Adjusted (no. scores) | -3.187 | 0.368 | 0.364 | 1.000 |
| Random score | Unadjusted | 0.001 | 0.391 | 0.387 | 0.051 |
| | Linear | 0.008 | 0.381 | 0.388 | 0.047 |
| | splines (3 DF) | 0.002 | 0.385 | 0.388 | 0.049 |
| Score closest to 12 months | Unadjusted | 0.000 | 0.394 | 0.388 | 0.056 |
| | Linear | 0.007 | 0.454 | 0.452 | 0.051 |
| | splines (3 DF) | -0.003 | 0.513 | 0.509 | 0.051 |

*Weighted by the number of follow-up scores, to account for difference in precision of mean scores with more follow-up scores

We next examined methods with regards to how they adjust for time since baseline, to illustrate the potential for bias if this trend is not modeled correctly (**Table S2**). For this set of results, we focus on the setting that assumes no intervention effect, and we examine both scenarios for how the intervention affects the follow-up time distribution to highlight how differential follow-up across arms could impact the performance. Under large linear or quadratic time trends in the outcome, we see substantial bias and elevated type 1 error rates under the optimal follow-up scenario when using the score closest to 12 months if the analysis does not adjust for the outcome timing. In the realistic follow-up scenario, we see some bias in the unadjusted model for both the randomly selected score and the score closest to 12 months. Under a quadratic trend in the outcome over time, in the scenarios we considered, the models that only adjust for time linearly had no or minimal residual bias or elevated type 1 error. However, given the potential for bias if a nonlinear relationship is present, and because additional adjustment does not noticeably impact efficiency of estimates under the realistic follow-up scenario (i.e., the empirical SE is very similar under linear vs. quadratic adjustment), adjusting for time flexibly (e.g., using natural cubic splines) is likely worthwhile to protect against a potentially nonlinear trend.

**Table S2.** Impact of adjustment for the timing of the follow-up score when using a random-selected score or the score closest to 12 months, under the setting of no intervention effect, for both the realistic and optimal scenario of follow-up for intervention participants, and different scenarios of the outcome trend over time.

| Modeling and Estimation Approach | | Realistic Follow-Up Scenario | | | | Optimal Follow-Up Scenario | | | |
|---|---|---|---|---|---|---|---|---|---|
| Choice of Follow-up Score | Adjustment for time since baseline | Bias | Empirical SE | SE (median) | Type 1 error | Bias | Empirical SE | SE (median) | Type 1 error |
| *Small trend in outcome over time* | | | | | | | | | |
| Random score | unadjusted | 0.014 | 0.388 | 0.388 | 0.051 | -0.005 | 0.391 | 0.387 | 0.051 |
| | linear | -0.002 | 0.387 | 0.390 | 0.048 | 0.008 | 0.381 | 0.388 | 0.047 |
| | splines (3 DF) | 0.009 | 0.393 | 0.392 | 0.051 | 0.002 | 0.385 | 0.388 | 0.049 |
| Score closest to 12 months | unadjusted | -0.013 | 0.390 | 0.388 | 0.051 | -0.075 | 0.394 | 0.388 | 0.062 |
| | linear | 0.003 | 0.386 | 0.389 | 0.048 | 0.007 | 0.454 | 0.452 | 0.051 |
| | splines (3 DF) | 0.009 | 0.389 | 0.390 | 0.052 | -0.003 | 0.513 | 0.509 | 0.051 |
| *Large trend in outcome over time* | | | | | | | | | |
| Random score | unadjusted | 0.213 | 0.394 | 0.393 | 0.084 | -0.053 | 0.395 | 0.393 | 0.056 |
| | linear | -0.002 | 0.387 | 0.390 | 0.048 | 0.008 | 0.381 | 0.388 | 0.047 |
| | splines (3 DF) | 0.009 | 0.393 | 0.392 | 0.051 | 0.002 | 0.385 | 0.388 | 0.049 |
| Score closest to 12 months | unadjusted | -0.149 | 0.395 | 0.392 | 0.071 | -0.756 | 0.397 | 0.390 | 0.493 |
| | linear | 0.003 | 0.386 | 0.389 | 0.048 | 0.007 | 0.454 | 0.452 | 0.051 |
| | splines (3 DF) | 0.009 | 0.389 | 0.390 | 0.052 | -0.003 | 0.513 | 0.509 | 0.051 |
| *Quadratic trend in outcome over time* | | | | | | | | | |
| Random score | unadjusted | 0.230 | 0.399 | 0.399 | 0.090 | -0.024 | 0.400 | 0.398 | 0.050 |
| | linear | -0.062 | 0.388 | 0.391 | 0.051 | 0.060 | 0.381 | 0.389 | 0.052 |
| | splines (3 DF) | 0.010 | 0.393 | 0.392 | 0.051 | 0.004 | 0.385 | 0.388 | 0.049 |
| Score closest to 12 months | unadjusted | -0.253 | 0.404 | 0.400 | 0.105 | -1.285 | 0.402 | 0.394 | 0.899 |
| | linear | -0.019 | 0.387 | 0.390 | 0.050 | -0.104 | 0.454 | 0.452 | 0.060 |
| | splines (3 DF) | 0.011 | 0.389 | 0.390 | 0.052 | 0.012 | 0.513 | 0.509 | 0.052 |

## *Sensitivity simulation scenario generating correlated data from an unstructured covariance*

To examine the sensitivity of our primary simulation study results on choice of correlations structure for the outcome generating model (exponential), we additionally examined a general (unstructured) correlation. The parameters in the unstructured correlation matrix were specified based on fitting an unstructured LMM to our retrospective cohort. Due to the fine temporal patterning of our PHQ outcome data that can be collected daily, to estimate unstructured correlation matrix we first grouped the daily time points into months and then replaced the original daily PHQ score with the average score of each patient within a month (if patients had multiple measures within a given month). The other parameters from this sensitivity simulation followed the scenario with a time-varying intervention effect, realistic scenario of follow-up for intervention participants, and assuming a small, linear outcome trend over time. The results are presented in **Table S3**.

**Table S3.** Sensitivity simulation results using an unstructured correlation matrix to simulate data

| Choice of Follow-up Score | Modeling time-varying intervention effect* | Mean Est. | Empirical SE | SE (median) | Coverage (95% CI) | Power |
|---|---|---|---|---|---|---|
| *Time-varying intervention effect* $(\Delta(12) = -1.5)$** | | | | | | |
| Score closest to 12 months | none | -0.989 | 0.507 | 0.512 | 0.829 | 0.489 |
| | splines (3DF) | -1.511 | 0.737 | 0.724 | 0.943 | 0.547 |
| Weighted GEE | none | -0.598 | 0.662 | 0.623 | ** | 0.171 |
| | splines (3DF) | -1.515 | 0.870 | 0.804 | 0.942 | 0.464 |
| LMM / Exchangeable | none | -0.587 | 0.469 | 0.465 | ** | 0.241 |
| | splines (3 DF) | -1.510 | 0.633 | 0.542 | 0.912 | 0.759 |
| LMM / CAR1 | none | -0.587 | 0.475 | 0.462 | ** | 0.254 |
| | splines (3 DF) | -1.505 | 0.585 | 0.586 | 0.948 | 0.728 |
| LMM / EXP | none | -0.596 | 0.476 | 0.462 | ** | 0.261 |
| | splines (3 DF) | -1.509 | 0.586 | 0.587 | 0.947 | 0.729 |

* Models estimating a time-varying effect report the estimated intervention effect at 12 months

**Models estimating a constant effect are estimating a different estimand (i.e., treatment effect) than the effect at 12 months. Under a time-varying intervention effect scenario, the value of this estimand depends on the distribution of the follow-up times and was empirically calculated as -0.997 when using the distribution of the score closest to 12 months. When using all scores, the true value of this estimand is not easily computable; we therefore do not present coverage.

As compared to the original exponential correlation structure (results in bottom rows of **Table 1** of the main paper), the power for all approaches decreased which was expected due to misspecification of the correlation structure. The linear mixed model with exchangeable covariance shows higher power than other methods. However, the approach may be unreliable when considering the discrepancy between the empirical and model standard errors. The coverage of 95% confidence interval is also below the nominal level. The LMM for time-varying effect that assumes exponential covariance showed consistent performance with higher power compared to other models with incorrect within-group correlation structures specified.